\documentclass[9pt,psfig]{article}
\usepackage{epsfig}
\usepackage[dvips]{color}
\begin{document}

\def\be{\begin{equation}}
\def\ee{\end{equation}}
\def\bq{\begin{equation}}
\def\eq{\end{equation}}
\def\bqa{\begin{eqnarray}}
\def\eqa{\end{eqnarray}}
\def\roughly#1{\mathrel{\raise.3ex
\hbox{$#1$\kern-.75em\lower1ex\hbox{$\sim$}}}}
\def\lsim{\roughly<}
\def\gsim{\roughly>}
\def\llgm{\left\lgroup\matrix}
\def\rrgm{\right\rgroup}
\def\vectrl #1{\buildrel\leftrightarrow \over #1}
\def\partrl{\vectrl{\partial}}
\def\gslash#1{\slash\hspace*{-0.20cm}#1}

\begin{center}
{\LARGE\bf
On the Color Dipole Picture}
\footnote{Presented at Diffraction 2016, Acireale (Catania, Sicily) September
2-8, 2016, AIP Conference Proceedings, ed. by A. Papa , to be published}
\end{center}
\begin{center}
{\bf Dieter Schildknecht$^{1,2,a)}$} \\[2.5mm]
{\small\sl $^1$Universit\"{a}t Bielefeld, Fakult\"{a}t f\"{u}r Physik 
Universit{\"a}tsstra{\ss}e 25, D-33615 Bielefeld} \\ 
{\small\sl $^2$Max-Planck-Institut f\"ur Physik 
F\"ohringer Ring 6, D-80805 M\"unchen} \\[0.3cm]
$^{a)}$Corresponding author: schild@physik.uni-bielefeld.de \\
URL: http://www.physik.uni-bielefeld.de
\end{center}

\vspace{0.3 cm}

\noindent{\small{\bf Abstract.}
We give a brief representation of the theoretical results from the color dipole
picture, covering the total photoabsorption cross section, high-energy $J/\psi$
photoproduction with respect to recent experimental data from the LHCb
Collaboration at CERN, and ultra-high energy neutrino scattering, relevant for
the ICE-CUBE experiment.}\\

\centerline{\bf DEEP INELASTIC SCATTERING}\medskip

\noindent
In terms of the imaginary part of the (virtual) forward Compton-scattering
amplitude, deep inelastic electron-proton scattering at low values of 
$x \cong Q^2/W^2 \lsim 0.1$ proceeds via $q \bar q$ forward scattering, compare
Figure 1.
\vspace*{0.2cm}

\begin{figure}[h]
\begin{minipage}[h]{7.5cm}
\epsfig{file=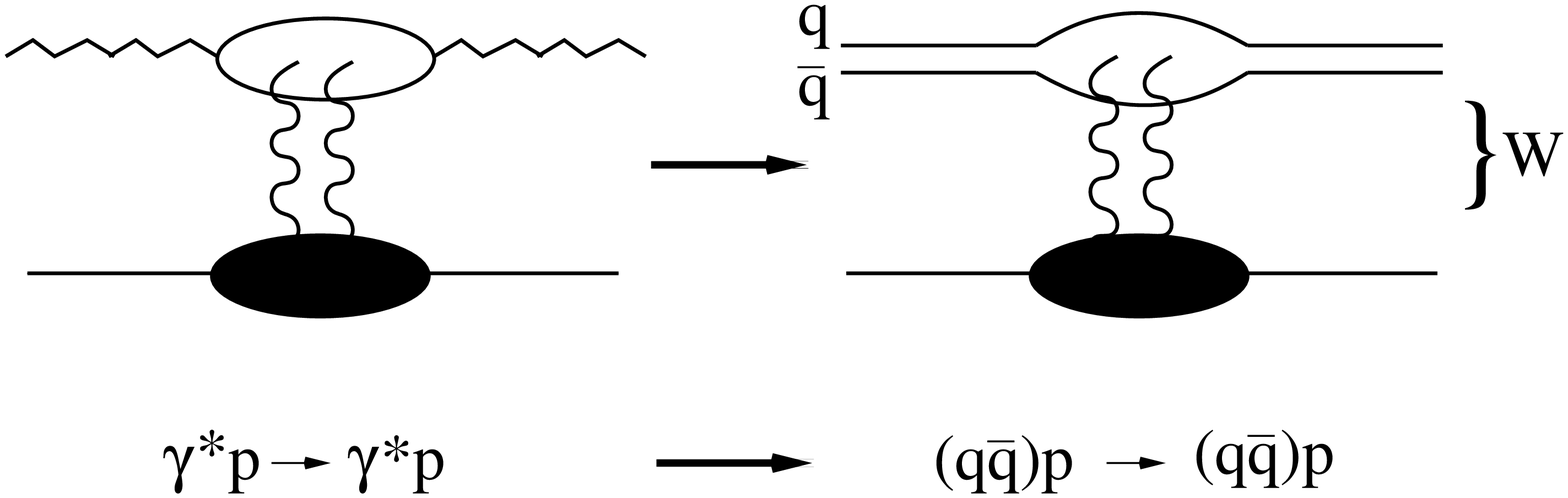,width=7cm}
\caption{
Relation between $\gamma^*p \to \gamma^*p$ and $(q \bar q)p 
\to (q \bar q)p$}
\label{Figure1}
\end{minipage}
\hspace*{1cm}
\begin{minipage}[h]{7.5cm}
\vspace*{-0.6cm}
\epsfig{file=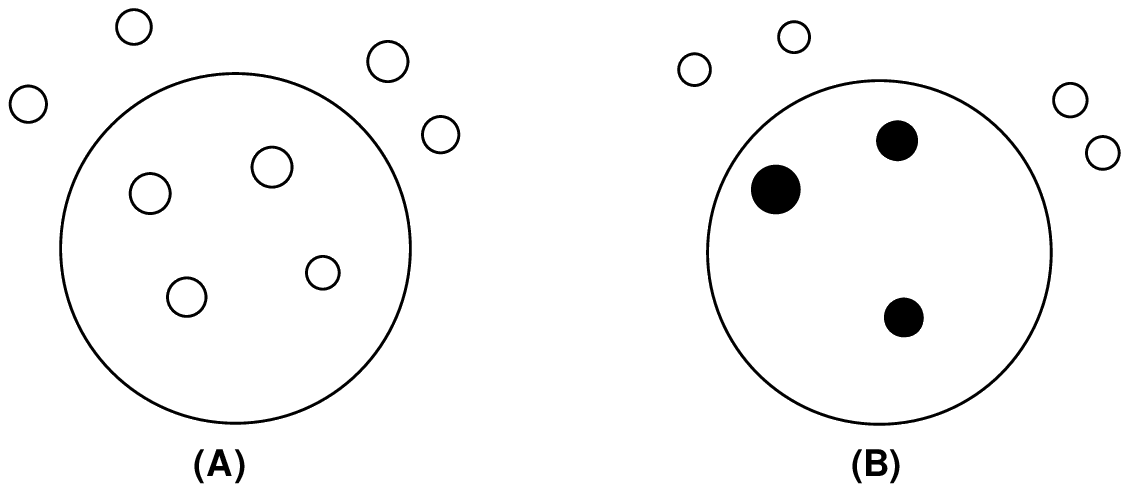,width=7cm}
\vspace*{-0.1cm}
\caption{
Color Transparency (A) and Saturation (B)}
\label{Figure2}
\end{minipage}
\end{figure}
\vspace*{0.2cm}

The total photoabsorption cross section is determined by (color-dipole
picture, CDP) 
\be
\sigma_{\gamma^*_{T,L}p} (W^2 , Q^2) = \int dz \int d^2 r_\bot 
| \psi_{T,L} (r_\bot , z(1-z), Q^2) |^2 
\sigma_{(q \bar q)^{J=1}_{T,L}p}
(\vec r_\bot \sqrt{z(1-z)}, W^2). 
\label{1}
\ee
Compare e.g. refs. \cite{bib1} for a
presentation of the CDP and a list of literature.
In standard notation, $\vert \psi_{T,L} (r_\bot , z(1-z), Q^2) \vert^2$ denotes
the probability for the photon of virtuality $Q^2$ to couple to a $(q \bar
q)^{J=1}_{T,L}$
state specified by the transverse size $\vec r_\bot$ and the longitudinal
momentum partition $0 \le z \le 1$, and $\sigma_{(q \bar q)^{J=1}_{T,L}p} (\vec
r_\bot \sqrt{z(1-z)}, W^2)$ denotes the color-dipole-proton cross section at
the total $\gamma^*p$ center-of-mass energy $W$. The gauge-invariant two-gluon
coupling of the $q \bar q$ dipole in Figure 1  requires a representation
of
the $(q \bar q)$-color-dipole-proton cross section of the form
\be
\sigma_{(q \bar q)^{J=1}_{T,L}p} (\vec r_\bot \sqrt{z(1-z)}, W^2) = \int d^2 l^
\prime_\bot \bar\sigma_{(q \bar q)^{J=1}_{T,L}p} (\vec l^{~\prime 2}_\bot , 
W^2)(1 - e^{-\vec l^{~\prime}_\bot \cdot \vec r_\bot \sqrt{z(1-z)}}) .
\label{2}
\ee
Note the factorization in (\ref{1}) into the $Q^2$-dependent ``photon wave
function'', and the $W^2$-dependence $(q \bar q)p$ cross section. The photon
wave function is known from quantum electrodynamics. It implies that at
sufficiently large values of $Q^2$ only small dipoles of transverse size $\vec
r^{~2}_\bot \sim 1/Q^2$ contribute to the interaction.

Concerning the $\vec r^{~2}_\bot$ dependence of the dipole cross section in
(\ref{2}), with an energy-dependent upper limit, $\vec l^{~2}_\bot \le
\vec l^{~2}_{\bot Max} (W^2)$ in (\ref{2}), for any fixed dipole size
$\vec r^{~2}_\bot$, we either have
i) $\vec l^{~\prime 2}_{\bot Max} (W^2) \vec r^{~2}_\bot \ll 1$, implying
$\sigma_{(q \bar q)p} \sim \vec r^{~2}$, ~~(``color transparency''), or
ii) $\vec l^{~\prime 2}_{\bot Max} (W^2) \vec r^{~\prime 2}_\bot \gg 1$, 
implying $\sigma_{(q \bar q)p} \sim \sigma^{(\infty)} (W^2)$, ~~
(``saturation''), compare Figure 2.

Evaluation of the photoabsorption cross section (\ref{1}), upon inserting the
dipole cross section in the limits i) and ii), translates
color transparency (c.tr.) and saturation (sat.) into specified limits of
photoabsorption \cite{bib1,bib2}.
\be
\sigma_{\gamma^*p} (W^2,Q^2) = \sigma_{\gamma^* p} (\eta (W^2,Q^2))
\sim \sigma^{(\infty)} (W^2) \left\{ \begin{array}{l@{\quad,\quad}l}
\frac{1}{\eta (W^2,Q^2)} & ~~~~{\rm for}~~\eta (W^2,Q^2) \gg 1, ~~{\rm
  c. tr.} \\
\ln \frac{1}{\eta (W^2,Q^2)} & ~~~{\rm for}~~ \eta (W^2,Q^2) \ll 1, ~~{\rm
  sat.} 
\end{array} \right. 
\label{5}
\ee
With $\sigma^{(\infty)} (W^2) \approx const$, the photoabsorption cross section
only depends on the single low-$x$ scaling variable
$
\eta (W^2,Q^2) = (Q^2 + m^2_0)/(\Lambda^2_{sat} (W^2)).
$
The ``saturation scale'', $\Lambda^2_{sat} (W^2)$ is determined by the first
moment of $\bar \sigma_{(q \bar q)^{J=1}_{T,L}p} (\vec l^{~\prime 2}_\bot,
W^2)$ in (\ref{2}), and $m^2_0 \lsim m^2_\rho$ for light quarks is fixed by
quark-hadron duality. Actually, $\sigma^{(\infty)} \sim \ln W^2$, i.e.
logarithmic violation of $\eta$-scaling.

\begin{figure}[h]
\begin{minipage}[h]{7cm}
\hspace*{-1cm}
\epsfig{file=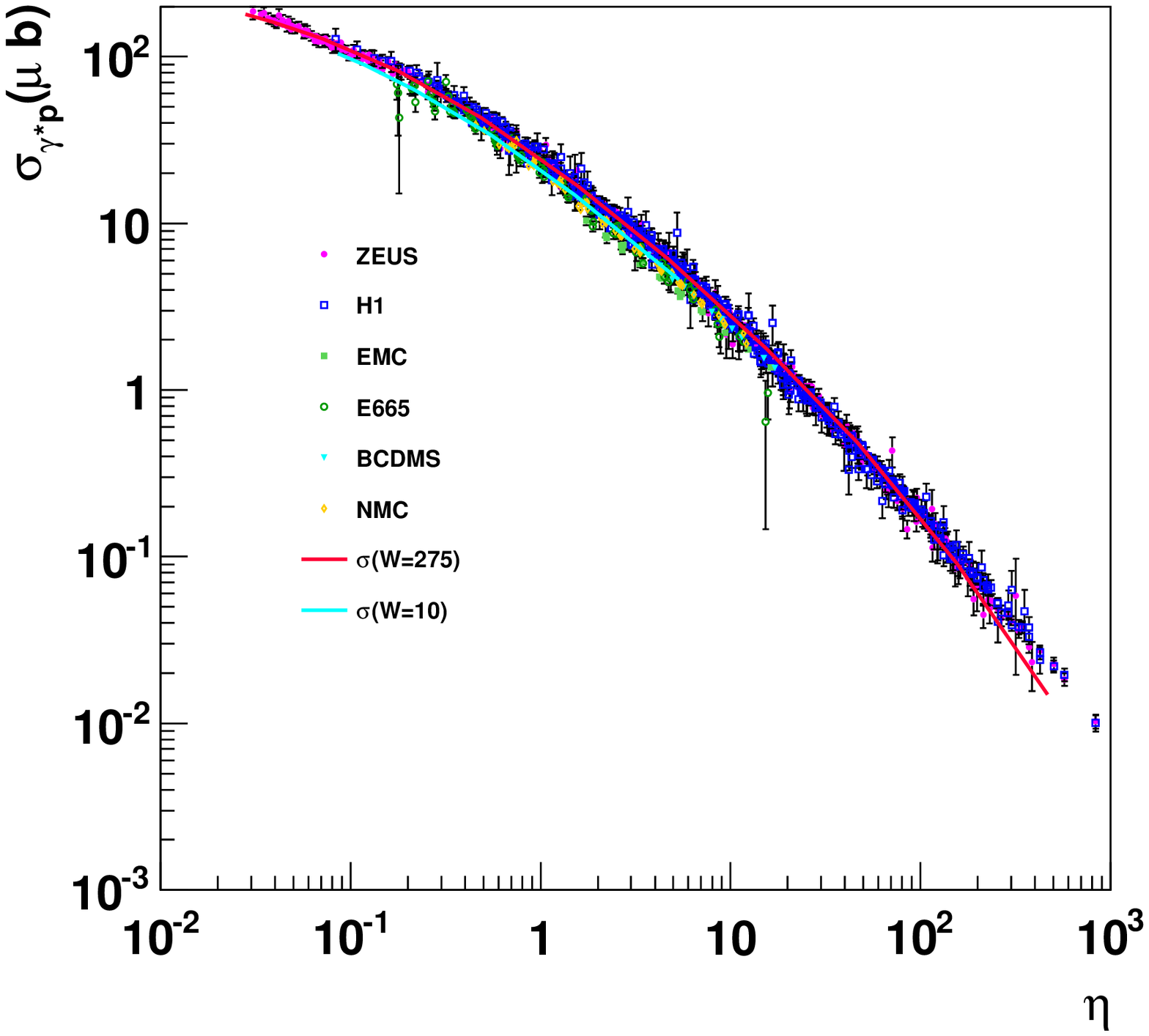,width=7cm}
\vspace*{-1.2cm}
\caption{
$\sigma_{\gamma^*p} (W^2,Q^2) = \sigma_{\gamma^*p}(\eta (W^2,Q^2))$.
}
\label{Figure3}
\end{minipage}
\begin{minipage}[h]{8.5cm}
\vspace*{0.4cm}
\epsfig{file=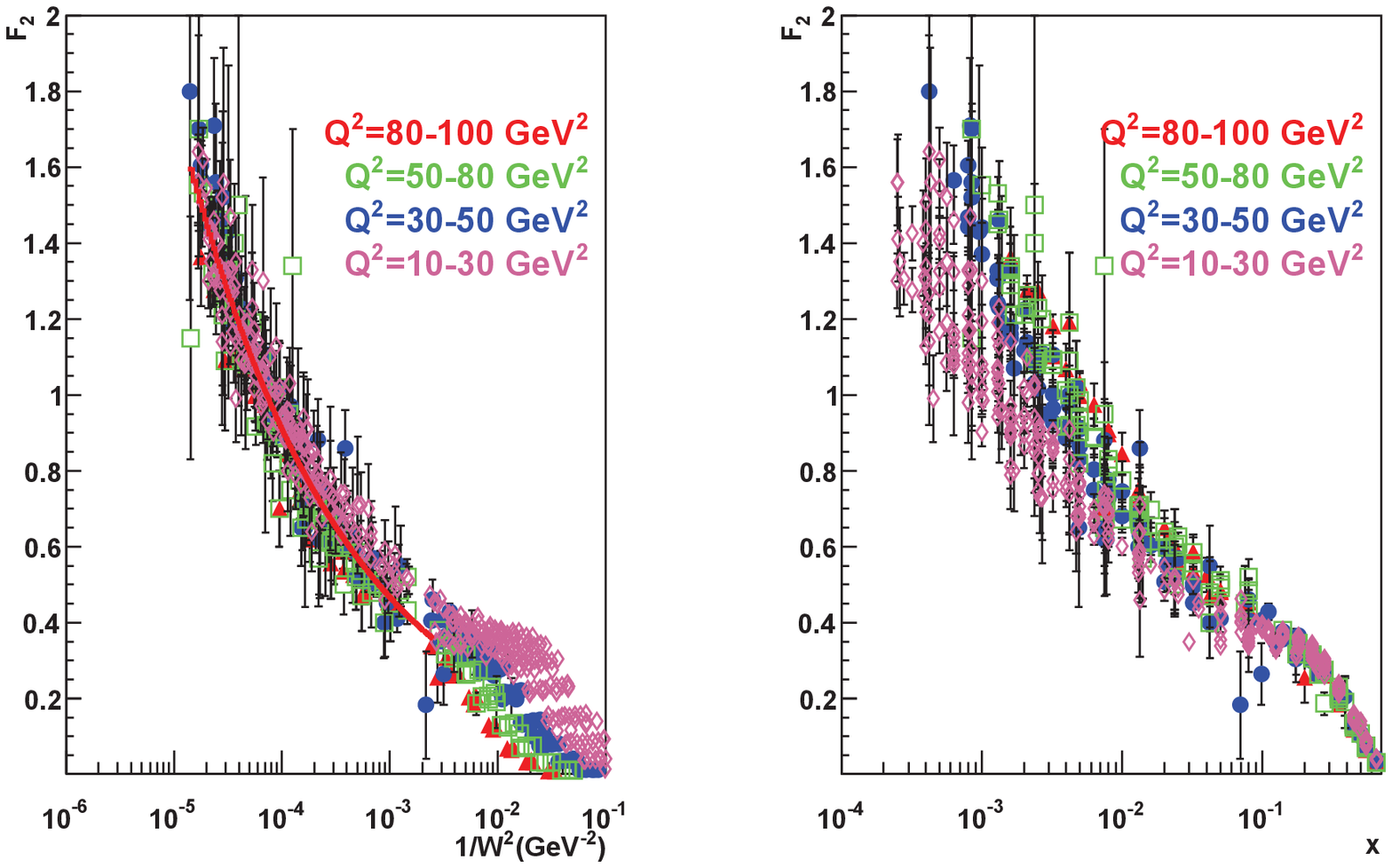,width=8.5cm}
\caption{
Proton structure function $F_2 (x,Q^2)$.
}
\label{Figure4}
\end{minipage}
\end{figure}

We summarize: the gauge-invariant two-gluon coupling of the color dipole
implies color transparency and saturation, which translate into
$\sigma_{\gamma^*p}\sim 1/\eta (W^2,Q^2)$ and $\sigma_{\gamma^*p} \sim \ln
(1/\eta (W^2,Q^2))$, respectively. No specific free-parameter-dependent ansatz
is necessary to arrive at this conclusion.
The experimental results agree with the above prediction, compare Figure 3. 
From (\ref{5}), for $W^2 \to \infty$ with $Q^2 > 0$ fixed, $\sigma_{\gamma^*p}
(\eta (W^2,Q^2))$ approaches \cite{bib2} the $Q^2=0$ photoproduction limit.
Compare Ref. \cite{bib1} for results on the longitudinal-to-transverse ratio.
In Figure 4, we show the proton structure function. In the color-transparency
region, as a consequence from $\sigma_{\gamma^*p} \sim 1/\eta (W^2,Q^2)$,
we have $F_2(x, Q^2) = F_2 (W^2)$.
\medskip

\centerline{\bf PHOTOPRODUCTION AND ELECTROPRODUCTION OF THE 
\boldmath$J/\psi$\unboldmath~ VECTOR MESON}

As depicted in Figure 5, the diffractive production of $q \bar q$ pairs,
in distinction from the total photoabsorption cross section in (\ref{1}),
depends on the square of the $(q \bar q)$-proton cross section.
Employing quark-hadron duality, diffractive vector-meson production is
obtained by integration of $q \bar q$ production over the mass interval that
it determined by the level spacing of the vector meson states under
consideration.
\vspace*{-0.8cm}
\begin{figure}[h]
\begin{center}
\epsfig{file=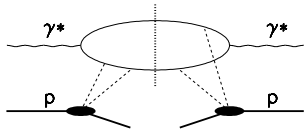, width=7cm}
\vspace*{-0.5cm}
\caption{Diffractive $\mathbf{q \bar q}$-pair production}
\end{center}
\label{Figure5}
\end{figure} 
\vspace*{-0.4cm}
In view of recent $J/\psi$-photoproduction data \cite{LHCb}
from the LHCb collaboration,
we concentrate \cite{Unique}
on $J/\psi$ production. Approximating $(c \bar c)$-production
by the cross section at the production threshold simplifies the integration 
over the level spacing $\Delta M^2_{J/\psi}$. This integration yields a
factor $\Delta F^2 (m^2_c, \Delta M^2{J/\psi})$, compare details in ref.
\cite{Ku-Schi}. Upon suppressing an overall constant factor, the $Q^2$
dependence and the $W^2$ dependence are given by \cite{Unique}
\be
\frac{d \sigma_{\gamma^* p \to J/\psi p} (W^2,Q^2)}{dt} \bigg|_{t \cong 0}
\propto \frac{(\sigma^{(\infty)} (W^2))^2}{\left(1 + \frac{Q^2+M^2_{J/\psi}}
{\Lambda^2_{sat} (W^2)}\right)^2}~~~ 
\frac{\Delta F^2(m^2_c, \Delta M^2_{J/\psi})}
{(Q^2 + M^2_{J/\psi})} 
\cong \left\{ \matrix{
\frac{\Lambda^4_{sat} (W^2) (\sigma^{(\infty)} (W^2))^2}{(Q^2 +
  M^2_{J/\psi})^2}~~~
\frac{\Delta F^2(m^2_c,~ \Delta M^2_{j/\psi})}{(Q^2 + M^2_{J/\psi})}\cr
\hspace*{-1cm}
(\sigma^{(\infty)} (W^2))^2~~~ \frac{\Delta F^2 (m^2_c, \Delta M^2_{J/\psi})}
{(Q^2 + M^2_{J/\psi})}}
\right. 
\label{10}
\ee
where the first line on the right-hand side refers to 
$\eta_{c \bar c} (W^2,Q^2) \equiv
(Q^2 + M^2_{J/\psi})/\Lambda^2_{sat} (W^2) \gg 1$, while the
second line refers to 
$\eta_{c \bar c} (W^2, Q^2) \equiv
(Q^2 + M^2_{J/\psi})/\Lambda^2_{sat} (W^2) \ll 1$.

At any fixed value of $Q^2$, for the $W$-dependence in (\ref{10}), we have
the significant limits of $\eta_{c \bar c} (W^2,Q^2) \gg 1$ and
$\eta_{c \bar c} (W^2,Q^2) \ll 1$, where
$\eta_{c \bar c} (W^2,Q^2) $
generalizes the low-$x$ scaling variable $\eta (W^2,Q^2)$.
The limits in (\ref{10}), respectively, correspond to color transparency
and saturation for $J/\psi$ production. Note that (\ref{10}) yields a
parameter-free prediction for $J/\psi$ production, once $\sigma^{(\infty)}
(W^2)$ and $\Lambda^2_{sat} (W^2)$ are known from the measurements of the
total photoabsorption cross section (\ref{1}).

For the comparison of the $Q^2$ dependence and the $W^2$ dependence in 
(\ref{10}) with the experimental data in the HERA energy range of
$W \lsim 300$ GeV, we refer to ref. \cite{Ku-Schi}. The figures in
ref. \cite{Ku-Schi} show good agreement of the prediction (\ref{10}) with
the experimental data.

Turning to photoproduction at $W \gsim 100$ GeV, from (\ref{10}), we obtain
\bqa
\sigma_{\gamma p \to J/\psi p} (W^2) & = & \frac{\left(1+
  \frac{M^2_{J/\psi}}{\Lambda^2_{sat} (W^2_1)}\right)^2}{\left(1+
\frac{M^2_{J/\psi}}{\Lambda^2_{sat} (W^2)} \right)^2}
\frac{\left(\sigma^{(\infty)} (W^2)\right)^2}{\left(\sigma^{(\infty)}
(W^2_1) \right)^2} \sigma_{\gamma p \to J/\psi p} (W^2_1 = (100~{\rm GeV})^2),
\nonumber \\
& \equiv & F_A (\Lambda^2_{sat} (W^2)) F_B (W^2) \sigma_{\gamma p \to
J/\psi p} (W^2_1 = (100~{\rm GeV})^2),
\label{12}
\eqa
where $\sigma_{\gamma p \to J/\psi} (W^2_1 = (100~{\rm GeV})^2) = 80 nb$ is to
be inserted on the right-hand side. From (\ref{12}), one finds the 
numerical results \cite{Unique} given in the last column of Table 1.
\begin{table}[h]
\caption{The $W$-dependence of $J/\psi$ photoproduction.}
\begin{tabular}{rlllll}
\hline
{\boldmath$W [{\rm GeV}]$} & {\boldmath$\Lambda^2_{sat} (W^2) [{\rm GeV}^2]$ }
& {\boldmath$\frac{M^2_{J/\psi}}{\Lambda^2_{sat} (W^2)}$}
& {\boldmath$F_A (\Lambda^2_{sat} (W^2))$} 
& {\boldmath$F_B (W^2)$} 
& {\boldmath$\sigma_{\gamma p \to J/\psi p} (W) [nb]$}  \\
\hline
100 & 4.32 & 2.22 & 1 & 1 & 80  \\
300 & 7.92 & 1.21 & 2.12 & 1.02 & 173 \\
1000 & 15.4 & 0.624 & 3.93 & 1.11 & 349 \\
2000 & 22.6 & 0.425 & 5.11 & 1.16 & 474 \\
\hline
\end{tabular}
\end{table}

\begin{figure}[h]
\begin{center}
\vspace*{-3cm}\hspace*{-8cm}\epsfig{file=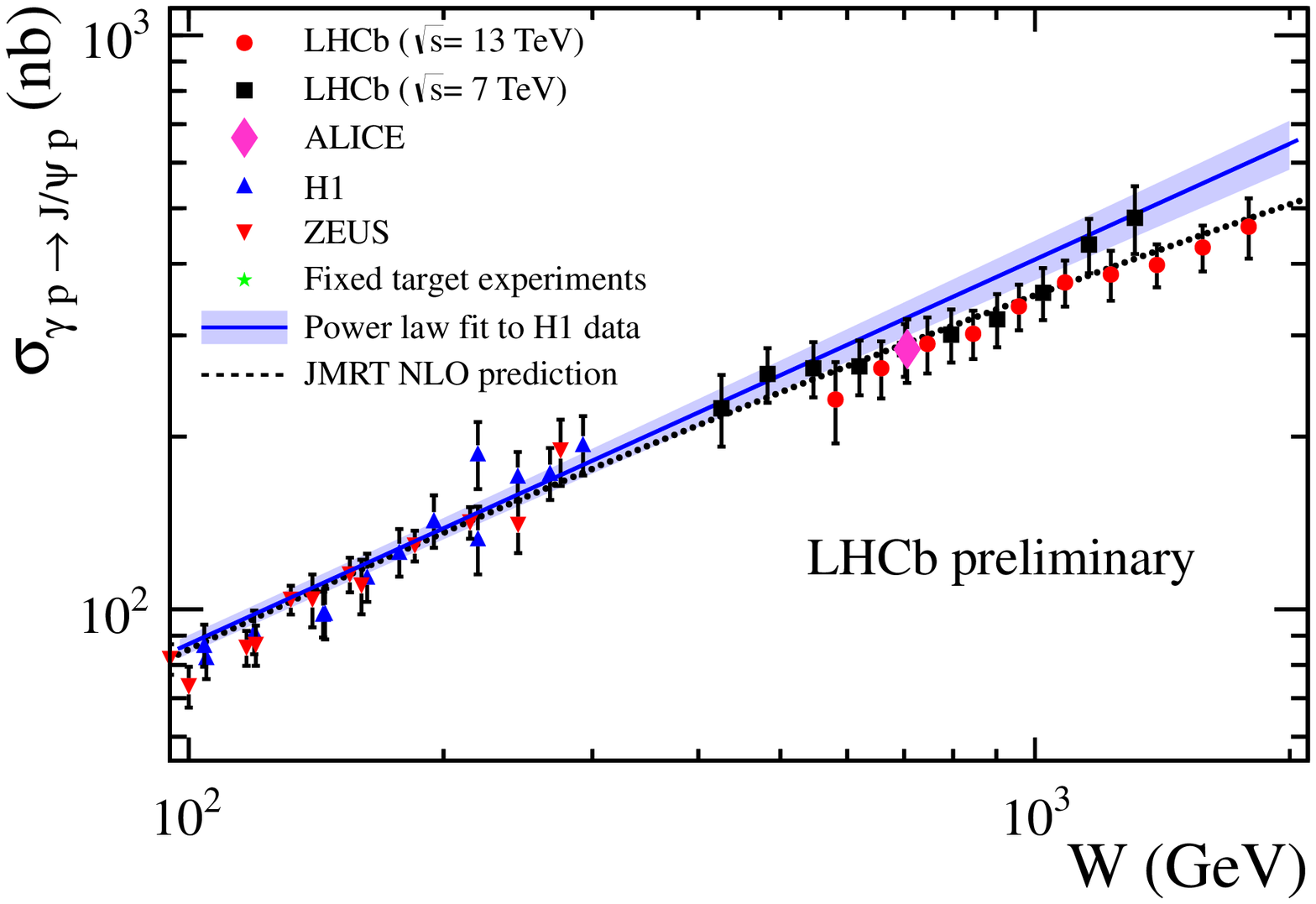, width=9cm}
\vspace*{-3cm}
\caption{The LHCB data (from ref. \cite{LHCb})}
\end{center}
\label{Figure6}
\end{figure} 
Comparison with the experimental data in Figure 6, taken from ref.
\cite{LHCb}, shows agreement with
our predictions in the last column of
Table 1. According to Table 1, the experimental data
from the LHCb collaboration, with $\eta_{c \bar c} (W^2,Q^2=0) = 
M^2_{J/\psi}/\Lambda^2_{sat} (W^2)$, lie in the range of $2.2 \gsim
\eta_{c \bar c} (W^2,Q^2) \gsim 0.43$. This region of $\eta_{c \bar c}$,
according to (\ref{10}), covers the transition from color transparency
to saturation; the deviation from the power-law fit in Figure 6 is to be
interpreted as a transition from color transparency to saturation. We
stress that the frequently assumed proportionality of $J/\psi$ photoproduction
to the square of the gluon distribution, corresponding to
$\Lambda^4_{sat} (W^2) \sim (\alpha_s (Q^2) x g (x, Q^2))^2$, violates
the necessary decent (logarithmic) high-energy saturation behavior that
is naturally contained in our prediction from the CDP.\medskip

\centerline{\bf THE NEUTRINO-NUCLEON CROSS SECTION AT ULTRA-HIGH ENERGIES}

\vspace*{0.2cm}

Predictions of the neutrino-nucleon cross section at ultrahigh energies
of the order of $E = 10^{6}~ {\rm GeV}$ 
and beyond are relevant and important for the 
interpretation of the search for cosmic neutrinos in e.g. the ICE-CUBE
experiment. Due to the presence of the $W$-boson mass in conjunction 
with ultrahigh energy, the process is determined by $x \simeq Q^2/W^2
\ll 0.1$. For the flavor-independent interaction of $q \bar q$ pairs, the
neutrino cross section is related to $\sigma_{\gamma^*p} (\eta (W^2,Q^2))$
by \cite{PRD88}
\be
\sigma_{\nu N}(E) = \frac{G^2_F M^4_W}{8 \pi^3 \alpha} 
\frac{n_f}{\sum_q Q^2_q} \int^{s-M_p^2}_{Q^2_{Min}} d Q^2 
\frac{Q^2}{(Q^2 + M^2_W)^2}
\int^{s-Q^2}_{M_p^2} \frac{dW^2}{W^2} \frac{1}{2} (1 + (1-y)^2) 
\sigma_{\gamma^*p} (\eta (W^2, Q^2)).
\label{13}
\ee
where $n_f/\sum_q Q^2_1 = 5/18$ for $n_f=4$ quark flavors. A careful
evaluation of the cross section shows that the cross section, even at
energies of the order of $E \sim 10^{10}~{\rm GeV}$ is dominated by the
color-transparency region, where $\sigma_{\gamma^*p} \sim 1/\eta (W^2,Q^2)$.
\vspace*{-1cm}
\begin{figure}[h]
\begin{minipage}[h]{7.5cm}
\epsfig{file=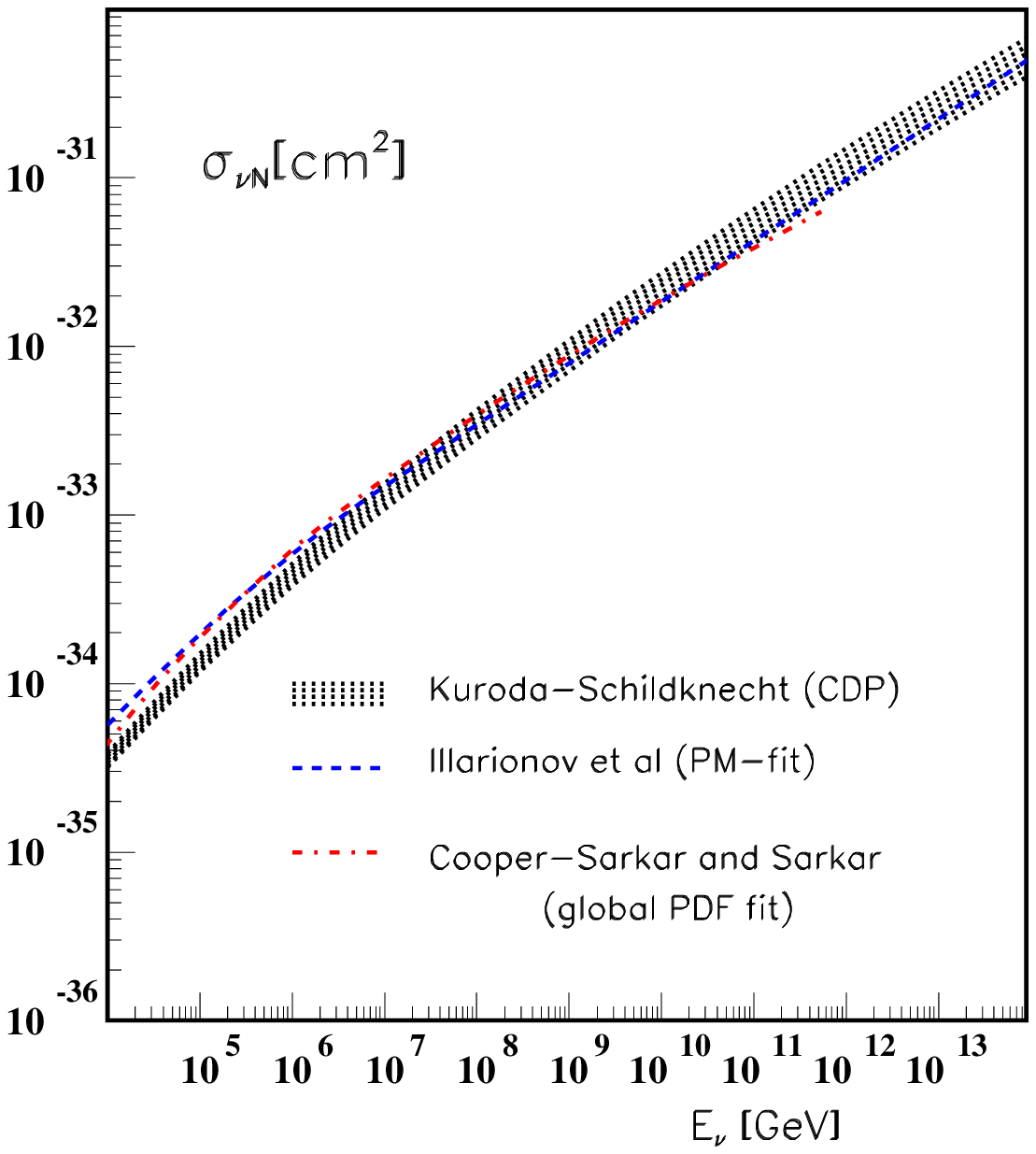,width=7cm}
\caption{
The neutrino-nucleon cross section \cite{PRD88}
compared with PDF fits.
}
\label{Figure7}
\end{minipage}
\hspace*{0.5cm}
\begin{minipage}[h]{7.5cm}
\epsfig{file=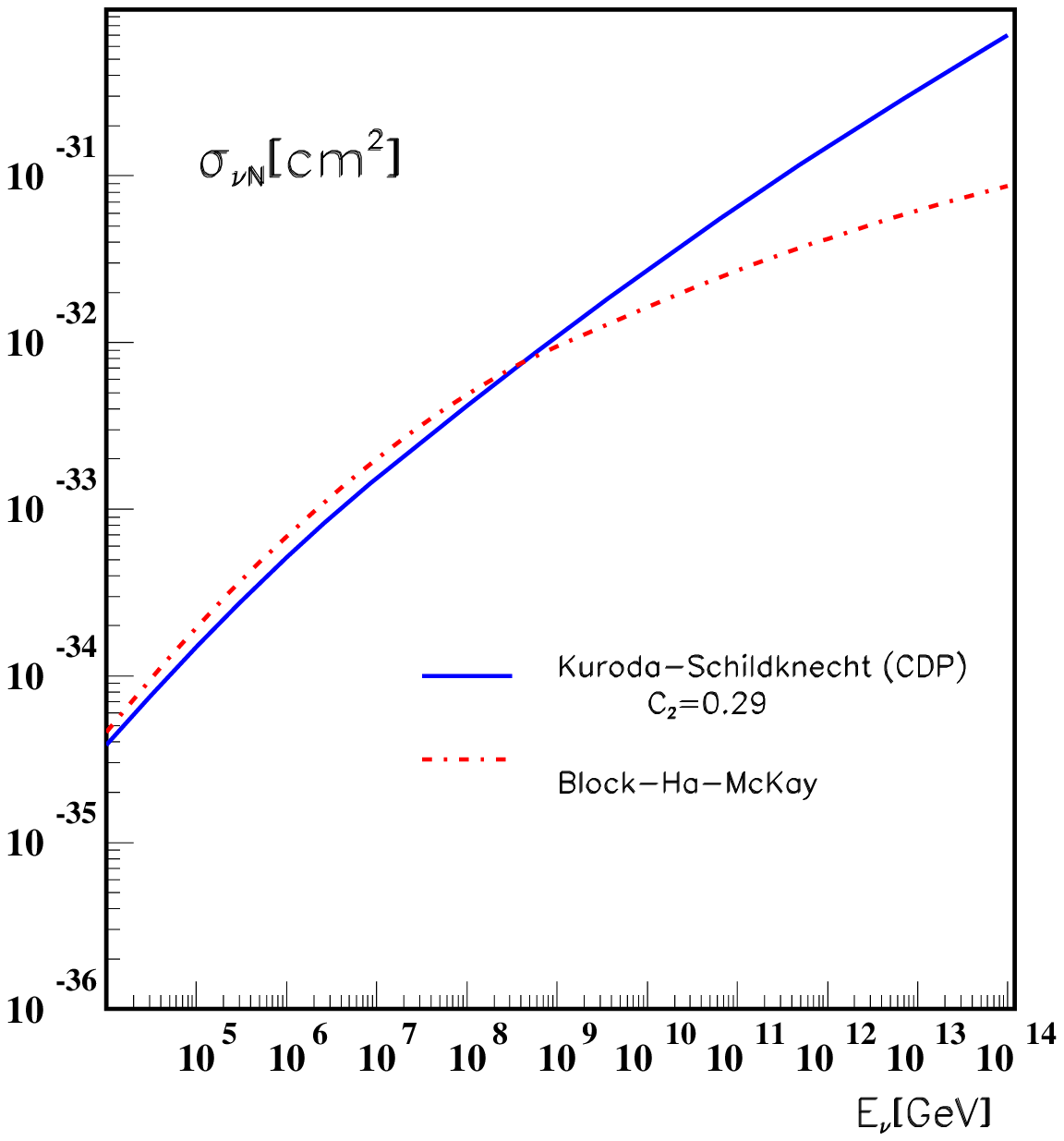,width=7cm}
\caption{
The neutrino-nucleon cross section and the 
Froissart-inspired result.
}
\label{Figure8}
\end{minipage}
\end{figure}


The simple explicit expression for the photoabsorption cross section, upon 
substitution into (\ref{13}) yields the results \cite{PRD88} in Figure 7. 
The extrapolation to ultrahigh energies in the CDP agrees with results
from the perturbative-QCD improved parton model. It is interesting to
note that the ``Froissart-inspired'' representation \cite{Block}
of the available
experimental data agrees with the results from the CDP
below $E_\nu \simeq 10^{10}~{\rm GeV}$, but yields a suppression of the cross
section for $E_\nu \gsim 10^{10}~{\rm GeV}$, compare Figure 8.




\end{document}